\documentclass[11pt,USenglish]{article}
\usepackage{amsmath}
\usepackage{amssymb}
\usepackage{amsthm}
\usepackage[english]{babel}
\usepackage{fullpage}
\usepackage[colorlinks=true,linkcolor=black,citecolor=black]{hyperref}
\usepackage{xcolor}

\usepackage{color}

\definecolor{darkgreen}{rgb}{0,.5,.2}

{\vspace{5mm}\begin{sloppypar}
  \noindent\hrulefill\, BEGIN \,\hrulefill
  \end{sloppypar}
}%
{\begin{sloppypar}
  \noindent\hrulefill\, END   \,\hrulefill
  \end{sloppypar}\vspace{5mm}
}

\newcounter{todo}

\newcounter{question}

\newcounter{comment}

\newcommand{\enumA}{(i)}
\newcommand{\enumB}{(ii)}
\newcommand{\enumC}{(iii)}

\theoremstyle{plain}

\newtheorem{example}{Example}

\theoremstyle{definition}
\newtheorem{remark}{Remark}

\newtheoremstyle{mydefinition}
	{4pt} 
	{-8pt} 
	{} 
	{} 
	{\bfseries} 
	{.} 
	{.5em} 
	{} 

\theoremstyle{mydefinition}
\newtheorem{defn}{Definition}
\newenvironment{definition}%
	{\textcolor{black!30}{\hrule}\begin{defn}}%
	{\end{defn}\textcolor{black!30}{\hrule}\bigskip}

\newcommand{\definedTerm}[1]{\textbf{#1}}

\newenvironment{grammar}
               {\par\begin{minipage}{\linewidth}\vspace*{3mm}\begin{footnotesize}\begin{textsf}}
               {\end{textsf}\end{footnotesize}\vspace*{3mm}\end{minipage}}

\newcommand{\mathSmallRDF}[1]{{\normalfont\texttt{\small #1}}}

\newcommand{\definedAs}    
	{=}

\newcommand{\fctsymDom}{\mathrm{dom}} 
\newcommand{\fctDom}[1]{\fctsymDom(#1)}

\newcommand{\tuple}[1]
	{\left( #1 \right)}

\newcommand{\symAllIRIs}{\mathcal{I}} 
\newcommand{\symAllLiterals}{\mathcal{L}} 
\newcommand{\symAllBNodes}{\mathcal{B}} 
\newcommand{\symAllTriples}{\mathcal{T}} 
\newcommand{\symAllVariables}{\mathcal{V}} 

\newcommand{\symAllTPs}{\mathcal{TP}} 
\newcommand{\symTP}{tp} 
\newcommand{\symBGP}{B} 
\newcommand{\symPattern}{P} 
\newcommand{\OpAND}{\text{ \normalfont\scriptsize\textsf{AND} }}
\newcommand{\OpUNION}{\text{ \normalfont\scriptsize\textsf{UNION} }}
\newcommand{\OpOPT}{\text{ \normalfont\scriptsize\textsf{OPT} }}
\newcommand{\OpFILTER}{\text{ \normalfont\scriptsize\textsf{FILTER} }}

\newcommand{\compatible}{\sim} 
\newcommand{\incompatible}{\not\sim} 

\newcommand{\fctsymCard}{\mathit{card}}
\newcommand{\fctCard}[1]{\fctsymCard(#1)}

\newcommand{\mcup}{\Cup}    
\newcommand{\msetminus}{{\setminus\!\!\setminus}\,}    

\def\ojoin{\setbox0=\hbox{$\Join$}\rule[.05ex]{.25em}{.45pt}\llap{\rule[1.1ex]{.25em}{.4pt}}}
\def\LJoin{\mathbin{\ojoin\mkern-7.5mu\Join}}

\newcommand{\fctEval}[2]{[\![#1]\!]_{#2}} 

\newcommand{\symAllStarTriples}{\mathcal{T}^{\!\star\!}} 
\newcommand{\symRDFstarGraph}{G^{\star}} 

\newcommand{\fctsymTermsPlus}{\mathrm{Elmts\!}^+} 
\newcommand{\fctTermsPlus}[1]{\fctsymTermsPlus\!(#1)} 

\newcommand{\fctsymTRefs}{\mathrm{Emb\!}^+} 
\newcommand{\fctTRefs}[1]{\fctsymTRefs\!(#1)} 

\newcommand{\symAllStarTPs}{\symAllTPs^{\star\!}} 
\newcommand{\OpAS}{\text{ \normalfont\scriptsize\textsf{AS} }}

\newcommand{\xxxSTAR}
	{$^\star$}

\newcommand{\rdfStar}{RDF\xxxSTAR}
\newcommand{\turtleStar}{Turtle\xxxSTAR}
\newcommand{\sparqlStar}{SPARQL\!\xxxSTAR}

\newcommand{\rdfStarTriple}{{\rdfStar}\! triple}
\newcommand{\rdfStarGraph}{{\rdfStar}\! graph}
\newcommand{\tpStar}{tri\-ple\xxxSTAR\! pattern}
\newcommand{\bgpStar}{BGP\xxxSTAR}
\newcommand{\solmapStar}{solution\xxxSTAR\! mapping}

\newcommand{\tRef}%
	{tref}

\title{Foundations of an Alternative Approach to Reification in RDF}
\author{Olaf Hartig \hspace{30mm} Bryan Thompson \\ {\normalsize Link\"oping University \hspace{20mm} Amazon Web Services \hspace{3mm}} \vspace{1mm} \\ {\normalsize http://olafhartig.de \hspace{65mm} ~}}

\begin{document}

\maketitle

\noindent
{\color{red}
This document has become obsolete. The ideas and formalizations in this document have been picked up and developed further by a task force\footnote{\url{https://w3c.github.io/rdf-star/}} of the RDF-DEV community group\footnote{\url{https://www.w3.org/community/rdf-dev/}} at the World Wide Web consortium (W3C). In this context, the approach has been renamed to RDF-star and SPARQL-star. The main result of the group is a community group report\footnote{\url{https://w3c.github.io/rdf-star/cg-spec}} that serves as a specification of the approach. Additionally, the group has developed a comprehensive collection of test suites\footnote{\url{https://w3c.github.io/rdf-star/tests/}} for different RDF-star syntaxes, for the RDF-star semantics, and for SPARQL-star. Notice that there are some aspects in which RDF-star and SPARQL-star, as defined in the aforementioned specification, differ from \rdfStar\ and \sparqlStar\ as defined in this document. Most notably, the semantics of embedded triples---which are now called quoted triples---is not anymore defined in terms of standard RDF reification and such a quoted triple is not anymore considered to be implicitly asserted in an RDF-star graph that contains nested triples of which the quoted triple is a constituent. Another difference is that the additional notion of BIND clauses as defined for \sparqlStar\ in this document~(cf.\ Example~\ref{ex:QueryingReificationDoneRight}) has not been carried over to SPARQL-star~(however, SPARQL-star has a new built-in function called TRIPLE which can be used for the same purpose).}

\medskip

\begin{abstract}
	\noindent
This document defines extensions of the RDF data model and of the SPARQL query language that
	capture an alternative approach to represent statement-level metadata.
While this alternative approach is backwards compatible with RDF reification as defined by the RDF standard, the approach aims to address usability and data management shortcomings of RDF~reification. 
One of the great advantages of the proposed approach is that it clarifies a means to \enumA~understand sparse matrices, the property graph model, hypergraphs, and other data structures with an emphasis on link attributes, \enumB~map such data onto RDF, and \enumC~query such data using SPARQL. Further, the proposal greatly expands both the freedom that database designers enjoy when creating physical indexing schemes and query plans for graph data annotated with link attributes and the interoperability of those database solutions.

\end{abstract}

\section{Introduction}

The RDF standard introduces the notion of reification as an approach to provide a set of RDF triples that describe some other RDF triple~\cite{Hayes14:RDFSemantics}. This form of state\-ment-level metadata about a reified triple has to include four additional RDF triples to refer to the reified triple.

\begin{example} \label{ex:Reification}
	Consider the following two RDF triples---given in Turtle syntax~\cite{PrudHommeaux14:Turtle}---that indicate the age of somebody named Bob\,%
	\footnote{Prefix declarations are omitted in all examples in this document. The prefixes used are the usual prefixes as can be found via the \url{http://prefix.cc} service.}
	\begin{verbatim}
	  :bob foaf:name "Bob" ; foaf:age 23 .
	\end{verbatim}
	To capture metadata about a given RDF triple as per RDF reification, we have to introduce an IRI or a blank node and use this IRI or blank node as the subject of four RDF triples that reify the given triple by applying the RDF reification vocabulary~\cite{Hayes14:RDFSemantics}. Then, the IRI or blank node can be used to provide metadata about the reified triple. For instance, by using a blank node, say {\normalfont\texttt{\_:s}}, we may reify the second of the two example triples as follows:
	\begin{verbatim}
	  _:s rdf:type rdf:Statement ;
	      rdf:subject :bob ;
	      rdf:predicate foaf:age ;
	      rdf:object 23 .
	\end{verbatim}
	Now, we can use the blank node to provide metadata about the triple:
	\begin{verbatim}
	  _:s dct:creator <http://example.com/crawlers#c1> ;
	      dct:source <http://example.net/homepage-listing.html> .
	\end{verbatim}
\end{example}

\noindent
RDF reification as demonstrated in the example has two major shortcomings.
First, adding four reification triples for every reified triple is inefficient for exchanging as well as for managing RDF data that includes state\-ment-level metadata.
Second, writing queries to access state\-ment-level metadata is cumbersome because any meta\-da\-ta-related (sub)expression in a query has to be accompanied by another subexpression to match the corresponding four reification triples.

\begin{example} \label{ex:QueryingReification}
	Consider the data (including the metadata) from Example~\ref{ex:Reification}. To query this data we may use SPARQL, the standard query language for RDF~\cite{Harris13:SPARQL1_1Language}. For instance, if we are interested in the age of Bob, including the source of this information, we may write a SPARQL query such as the following.
	\begin{verbatim}
	  SELECT ?age ?src WHERE {
	     ?bob foaf:name "Bob" ;
	          foaf:age ?age .
	     ?r rdf:type rdf:Statement ;
	        rdf:subject ?b ;
	        rdf:predicate foaf:age ;
	        rdf:object ?age ;
	        dct:source ?src .
	  }
	\end{verbatim}
	Note that the given query contains four triple patterns to identify the reified triple whose metadata we want to see. If we were also interested in potential metadata about the corresponding {\normalfont\texttt{foaf:name}} triple, we would have to add another four, reification-related triple patterns.
\end{example}

\noindent
During a Dagstuhl seminar on ``Semantic Data Management''~\cite{Antoniou12:DagstuhlSeminar}
	several
participants of the seminar---including Bryan Thompson (Systap, LLC), Orri Erling and Yrj\"an\"a Rankka (OpenLink Software), and Olaf Hartig (then Humboldt Universit\"at zu Berlin)---discussed an alternative approach to reification that addresses the aforementioned shortcomings. 

This document provides a formal foundation for this approach.
The document is structured as follows: Section~\ref{sec:Informal} outlines the approach informally. Thereafter, Section~\ref{sec:RDFstar} introduces an extension of the RDF data model that makes metadata statements a first class citizen. Finally, Sections \ref{sec:SPARQLstar} and \ref{sec:W3CExt} extends the query language SPARQL to enable users to benefit from the extended data model.

\section{An Alternative Approach to Reification in RDF} \label{sec:Informal}

The alternative approach to reification is based on
	the idea of using a triple directly in the subject position or object position of (other) triples that represent metadata about the embedded triple.

\begin{example} \label{ex:ReificationDoneRight}
	Assume a possible extension of the Turtle syntax that implements the idea of embedding RDF triples into other RDF triples by enclosing any embedded triple in '\texttt{\footnotesize \textless\textless}' and '\texttt{\footnotesize \textgreater\textgreater}' (Section~\ref{ssec:RDFstar:TurtleStar} shall introduce such an extension). Then, the data from Example~\ref{ex:Reification} (including the metadata) could be represented as follows.
	\begin{verbatim}
	  :bob foaf:name "Bob" .
	  <<:bob foaf:age 23>> dct:creator <http://example.com/crawlers#c1> ;
	                       dct:source <http://example.net/homepage-listing.html> .
	\end{verbatim}
\end{example}

\noindent
Embedding triples into (metadata) triples as demonstrated in the example
	achieves a more compact representation of statement-level metadata than standard RDF reification.
Such a compact representation may improve comprehensibility for users who have to inspect RDF documents directly~(e.g., in a text editor). Such a representation may also reduce the size of RDF documents that include statement-level metadata and, thus, may be advantageous for data exchange.
Furthermore, embedded triples, conceived of as a form of self-referencing identifiers, corresponds naturally to the concept of triple identifiers that some RDF data management systems
	such as Systap's Bigdata~\cite{Thompson14:Bigdata}
use internally to avoid the overhead of keeping a physical representation of four reification triples per reified triple.

	Given that triples are embedded into other triples, the idea of such an embedding
can be carried over to SPARQL queries.

\begin{example} \label{ex:QueryingReificationDoneRight}
	By adopting the extended syntax outlined in Example~\ref{ex:ReificationDoneRight}, we could represent the query from Example~\ref{ex:QueryingReification}
		in the following, more compact form.
	\begin{verbatim}
	  SELECT ?age ?src WHERE {
	     ?bob foaf:name "Bob" .
	     <<?bob foaf:age ?age>> dct:source ?src .
	  }
	\end{verbatim}
	An alternative
		form is to use BIND clauses as follows.
\newpage 
	\begin{verbatim}
	  SELECT ?age ?src WHERE {
	     ?bob foaf:name "Bob" .
	     BIND( <<?bob foaf:age ?age>> AS ?t )
	     ?t dct:source ?src .
	  }
	\end{verbatim}
\end{example}

\noindent
The remainder of this document provides a formal definition of the approach outlined
	in this section.
An important characteristic of this formalization is its backward compatibility with standard RDF~reification.

\section{\rdfStar\ -- A Metadata Extension of RDF} \label{sec:RDFstar}
This section introduces an extension of the RDF data model~\cite{Cyganiak14:RDFConcepts} that makes metadata~statements a first class citizen. 
Hereafter, the extended data model is referred to as \rdfStar.

\subsection{Concepts}

Assume
pairwise disjoint sets $\symAllIRIs$ (all IRIs), $\symAllBNodes$ (blank nodes), and $\symAllLiterals$ (literals). As usual,
	an \emph{RDF triple} is a tuple $\tuple{s,p,o} \in (\symAllIRIs \cup \symAllBNodes) \times \symAllIRIs \times (\symAllIRIs \cup \symAllBNodes \cup \symAllLiterals)$ and an \emph{RDF graph} is a set of RDF~triples.


\rdfStar\ extends such triples by permitting the embedding of a given triple in the subject or object position of another triple. Triples whose subject or object is an embedded triple represent some form of metadata. An embedded triple may itself be a metadata triple and, thus, may also contain embedded triples; and so forth.
%
	The following definition captures this notion.

\begin{definition} \label{def:TripleStar}
	Let $\symAllStarTriples$ be an (infinite) set of tuples that is defined recursively as follows:
	\begin{enumerate}
		\item $\symAllStarTriples$ includes all RDF triples, i.e., $\symAllStarTriples \supseteq (\symAllIRIs \cup \symAllBNodes) \times \symAllIRIs \times (\symAllIRIs \cup \symAllBNodes \cup \symAllLiterals)$; and
		\item if $t \in \symAllStarTriples$ and $t' \in \symAllStarTriples$, then $\tuple{t,p,o} \in \symAllStarTriples$, $\tuple{s,p,t} \in \symAllStarTriples$ and $\tuple{t,p,t'} \in \symAllStarTriples$ for all $s \in ( \symAllIRIs \cup \symAllBNodes )$, $p \in \symAllIRIs$, and $o \in ( \symAllIRIs \cup \symAllBNodes \cup \symAllLiterals)$.
	\end{enumerate}
	Any tuple $\tuple{s,p,o} \in \symAllStarTriples$ is an \definedTerm{\rdfStarTriple}.
	A set of {\rdfStarTriple}s is called an \definedTerm{\rdfStarGraph}.
\end{definition}

\noindent
Hereafter, for any \rdfStarTriple\ $t \in \symAllStarTriples$\!, $\fctTermsPlus{t}$ denotes the set of all RDF terms and all {\rdfStarTriple}s mentioned in $t$; i.e., if $t = \tuple{s,p,o}$, then $\fctTermsPlus{t} \definedAs \lbrace s,p,o \rbrace \cup \big\lbrace x' \in \fctTermsPlus{x} \,\big|\, x \in \lbrace s,o \rbrace \cap \symAllStarTriples \big\rbrace$. An \rdfStarTriple\ $t$ with $\fctTermsPlus{t} \cap \symAllStarTriples \neq \emptyset$ is called a \emph{metadata triple} (note that any other \rdfStarTriple\ is an ordinary RDF triple).



Overloading function $\fctsymTermsPlus$, for any \rdfStarGraph\
	$\symRDFstarGraph$\!,
$\fctTermsPlus{\symRDFstarGraph} \definedAs \bigcup_{t \in \symRDFstarGraph} \fctTermsPlus{t}$. Furthermore, $\fctTRefs{\symRDFstarGraph}$ denotes the set of all {\rdfStarTriple}s that are (recursively) embedded in {\rdfStarTriple}s of \rdfStarGraph\ $\symRDFstarGraph$; i.e., $\fctTRefs{\symRDFstarGraph} \definedAs \fctTermsPlus{\symRDFstarGraph} \cap \symAllStarTriples$.

\begin{example} \label{ex:RDFstarGraph}
	The data represented in Example~\ref{ex:ReificationDoneRight} can be parsed into the following \rdfStarGraph.
	\begin{align*}
		\symRDFstarGraph_\mathsf{ex} = \big\lbrace
			& \tuple{\mathSmallRDF{:bob}, \, \mathSmallRDF{foaf:name}, \, \mathSmallRDF{"Bob"}} ,\\
			& \tuple{ \tuple{\mathSmallRDF{:bob}, \mathSmallRDF{foaf:age}, \mathSmallRDF{23}}, \, \mathSmallRDF{dct:creator}, \, \mathSmallRDF{http://example.com/crawlers\#c1}} ,\\
			&\tuple{ \tuple{\mathSmallRDF{:bob}, \mathSmallRDF{foaf:age}, \mathSmallRDF{23}}, \, \mathSmallRDF{dct:source}, \, \mathSmallRDF{http://example.net/homepage-listing.html}}
		\big\rbrace
	\end{align*}
	Hence, this \rdfStarGraph\ consists of three {\rdfStarTriple}s, and its set of embedded {\rdfStarTriple}s contains a single triple, that is, $\fctTRefs{\symRDFstarGraph_\mathsf{ex}} = \big\lbrace \tuple{\mathSmallRDF{:bob}, \mathSmallRDF{foaf:age}, \mathSmallRDF{23}} \big\rbrace$.
\end{example}

\subsection{\rdfStar\ Semantics} \label{ssec:RDFstarSemantics}
To
	support a model-theoretic interpretation of
{\rdfStarGraph}s in terms of the standard RDF semantics~\cite{Hayes14:RDFSemantics} this section defines a transformation from {\rdfStarGraph}s to ordinary RDF graphs. This transformation may also be used to enable ordinary RDF data management systems (that do not support \rdfStar) to process data that is represented as an \rdfStarGraph.

The transformation is based on the following three functions.
First, the transformation uses a function that associates every embedded \rdfStarTriple\ $t \in \fctTRefs{\symRDFstarGraph}$ in an \rdfStarGraph~$\symRDFstarGraph$ with a fresh and unique blank node. Hence, this function is called a \emph{bnode assignment function}. 

\begin{definition}
	A \definedTerm{bnode assignment function} $id$ for an \rdfStarGraph~$\symRDFstarGraph$ is a bijective function $id \!: \fctTRefs{\symRDFstarGraph} \rightarrow B$ such that $B \subseteq \symAllBNodes$ is a set of blank nodes
		that has the following two properties: \enumA~$\left| B \right| = \left| \fctTRefs{\symRDFstarGraph} \right|$ and \enumB~$B \cap \fctTermsPlus{\symRDFstarGraph} = \emptyset$.
\end{definition}

\noindent
Second, the transformation uses a \emph{reification function} that associates every embedded \rdfStarTriple\ in an \rdfStarGraph\ with a corresponding set of four reification triples.

\begin{definition}
	Let~$\symRDFstarGraph$ be an \rdfStarGraph\ and let~$id$ be a bnode assignment function for~$\symRDFstarGraph$\!. The $id$-spe\-cif\-ic \definedTerm{reification function} for~$\symRDFstarGraph$ is a function $\mathrm{reif}^{id} \!: \fctTRefs{\symRDFstarGraph} \rightarrow 2^{\symAllStarTriples}$ that, for every (embedded) \rdfStarTriple\ $t \in \fctTRefs{\symRDFstarGraph}$, is defined as~follows:
	\begin{align*}
		\mathrm{reif}^{id}\bigl( t \bigr) \definedAs \big\lbrace
			& \tuple{id^*\!(t),\mathSmallRDF{rdf\!:\!type},\mathSmallRDF{rdf\!:\!Statement}},
			\, \tuple{id^*\!(t),\mathSmallRDF{rdf\!:\!subject},id^*\!(s)}, \\
			& \tuple{id^*\!(t),\mathSmallRDF{rdf\!:\!predicate},id^*\!(p)},
			\, \tuple{id^*\!(t),\mathSmallRDF{rdf\!:\!object},id^*\!(o)}
		\big\rbrace ,
	\end{align*}
	where $id^*\!(t) \definedAs id(t)$ for all $t \in \fctTRefs{\symRDFstarGraph}$ and $id^*\!(x) \definedAs x$ for all $x \notin \fctTRefs{\symRDFstarGraph}$.
\end{definition}

\noindent
The third function for the transformation unfolds
	(potentially nested)
{\rdfStarTriple}s recursively. 

\begin{definition}
	Let~$\symRDFstarGraph$ be an \rdfStarGraph\ and let~$id$ be a bnode assignment function for~$\symRDFstarGraph$\!. The $id$-spe\-cif\-ic \definedTerm{unfold function} for~$\symRDFstarGraph$ is a function $\mathrm{rdf}^{id} \!: \bigl( \symRDFstarGraph \cup \fctTRefs{\symRDFstarGraph} \bigr) \rightarrow 2^{\symAllTriples}$ that, for every \rdfStarTriple\ $t \in \bigl( \symRDFstarGraph \cup \fctTRefs{\symRDFstarGraph} \bigr)$ with $t = \tuple{s,p,o}$, is defined as follows:

	\begin{equation*}
		\mathrm{rdf}^{id}(t) \definedAs \begin{cases}
			\big\lbrace \tuple{id(s),p,o} \big\rbrace \cup \mathrm{reif}^{id}(s) \cup \mathrm{rdf}^{id}(s) & \text{if $s \in \symAllStarTriples$ and $o \notin \symAllStarTriples$,} \\[1mm]
			\big\lbrace \tuple{s,p,id(o)} \big\rbrace \cup \mathrm{reif}^{id}(o) \cup \mathrm{rdf}^{id}(o) & \text{if $s \notin \symAllStarTriples$ and $o \in \symAllStarTriples$,} \\[1mm]
			\big\lbrace \tuple{id(s),p,id(o)} \big\rbrace \cup \mathrm{reif}^{id}(s) \cup \mathrm{rdf}^{id}(s) & \text{if $s \in \symAllStarTriples$ and $o \in \symAllStarTriples$,} \\[-1mm]
			\hspace{30mm} \cup \, \mathrm{reif}^{id}(o) \cup \mathrm{rdf}^{id}(o) & \\
			\big\lbrace \tuple{s,p,o} \big\rbrace & \text{else} .
		\end{cases}
\vspace{1mm}
	\end{equation*}
\end{definition}

\noindent
Given these three functions, the transformation itself is defined as follows.

\begin{definition} \label{def:Unfolding}
	Let~$\symRDFstarGraph$ be an \rdfStarGraph\ and let~$id$ be a bnode assignment function for~$\symRDFstarGraph$\!. The $id$-spe\-cif\-ic \definedTerm{unfolded RDF graph} of~$\symRDFstarGraph$\!, denoted by $\mathrm{rdf}^{id}(\symRDFstarGraph)$, is an RDF graph that is defined as~follows:
	\begin{equation*}
		\mathrm{rdf}^{id}(\symRDFstarGraph) \definedAs \bigcup_{t \in \symRDFstarGraph} \mathrm{rdf}^{id}(t) .
\vspace{1mm}
	\end{equation*}
\end{definition}

\begin{remark}
	Due to the definition of the unfold function, the transformation as given in Definition~\ref{def:Unfolding} entails any RDF triple $t \in \fctTRefs{\symRDFstarGraph}$ that is embedded in some metadata triple in an \rdfStarGraph~$\symRDFstarGraph$. Hence, the given transformation captures the use case of RDF reification in which
		RDF graphs that contain a reification of an RDF triple $t$ also contain $t$ itself.
\end{remark}

\begin{example} \label{ex:Unfolding}
	An unfolded RDF graph of the \rdfStarGraph\ in Example~\ref{ex:RDFstarGraph} is given as follows:
	\begin{align*}
		\mathrm{rdf}^{id_\mathsf{ex}}(\symRDFstarGraph_\mathsf{ex}) = \big\lbrace
			& \tuple{\mathSmallRDF{:bob}, \, \mathSmallRDF{foaf:name}, \, \mathSmallRDF{"Bob"}} ,\\
			& \tuple{\mathSmallRDF{:bob}, \, \mathSmallRDF{foaf:age}, \, \mathSmallRDF{23}} ,\\
			& \tuple{ b, \, \mathSmallRDF{rdf:type}, \, \mathSmallRDF{rdf:Statement} } ,\\
			& \tuple{ b, \, \mathSmallRDF{rdf:subject}, \, \mathSmallRDF{:bob} } ,\\
			& \tuple{ b, \, \mathSmallRDF{rdf:predicate}, \, \mathSmallRDF{foaf:age} } ,\\
			& \tuple{ b, \, \mathSmallRDF{rdf:object}, \, \mathSmallRDF{23} } ,\\
			& \tuple{ b, \, \mathSmallRDF{dct:creator}, \, \mathSmallRDF{http://example.com/crawlers\#c1} } ,\\
			&\tuple{ b, \, \mathSmallRDF{dct:source}, \, \mathSmallRDF{http://example.net/homepage-listing.html} }
		\big\rbrace
	\end{align*}
	Note that this example uses a bnode assignment function $id_\mathsf{ex}$ that associates the (embedded) \rdfStarTriple\ $\tuple{\mathSmallRDF{:bob}, \, \mathSmallRDF{foaf:age}, \, \mathSmallRDF{23}}$ with blank node $b \in \symAllBNodes$.
\end{example}

\subsection{\turtleStar\ -- An \rdfStar\ Extension of Turtle} \label{ssec:RDFstar:TurtleStar}
Example~\ref{ex:ReificationDoneRight} outlines a possible extension of the Turtle syntax to write an \rdfStarGraph. This section defines this extension, called \turtleStar.

\turtleStar\ extends the Turtle grammar (as given in~\cite[Section~6.5]{PrudHommeaux14:Turtle}) with the following three additional productions.
\begin{grammar}
tripleX  ::=  '\textless\textless' \, subjectX \, predicate \, objectX \, '\textgreater\textgreater	' \par\vspace{1mm}
subjectX ::=  iri $\vert$ BlankNode $\vert$ tripleX \par\vspace{1mm}
objectX ::=  iri $\vert$ BlankNode $\vert$ literal $\vert$ tripleX \par
\end{grammar}

\noindent
Any string that matches production \textsf{tripleX} is to be mapped to an \rdfStarTriple\ $\tuple{s,p,o}$ such that
\enumA~$s$ is the RDF term or the (embedded) \rdfStarTriple\ that can be obtained by parsing the substring that matches \textsf{subjectX},
\enumB~$p$ is the RDF term obtained by parsing the substring that matches \textsf{predicate}, and
\enumC~$o$ is the RDF term or the (embedded) \rdfStarTriple\ obtained by parsing the substring that matches \textsf{objectX}.

In addition to adding these three productions to the grammar, \turtleStar\ extends the productions labeled \texttt{[10]} and \texttt{[12]} in the standard Turtle grammar as follows (the extension to the productions are given in bold font).
\begin{grammar}
subject ::=  iri $\vert$ BlankNode $\vert$ collection \textbf{$\pmb{\vert}$ tripleX } \par\vspace{1mm}
object ::=  iri $\vert$ BlankNode $\vert$ collection $\vert$ blankNodePropertyList $\vert$ literal \textbf{$\pmb{\vert}$ tripleX }
\par
\end{grammar}

\noindent
A \turtleStar\ parser is a Turtle parser that is extended to take into account the productions defined in this section. Hence, such a parser constructs a set of {\rdfStarTriple}s (i.e., an \rdfStarGraph) that can be processed by an \rdfStar-aware system.

Note that ordinary RDF data management systems (that do not support \rdfStar) may easily be enabled to read a \turtleStar\ document and process the data; they only need to use a \turtleStar\ parser equipped with a transformation component that transforms the \rdfStarGraph\ given in the document to an unfolded RDF graph as defined in Section~\ref{ssec:RDFstarSemantics}.

\section{\sparqlStar\ -- A Metadata Extension of SPARQL} \label{sec:SPARQLstar}

This section introduces
	\sparqlStar, which is an \rdfStar-aware extension of the RDF query language SPARQL; i.e., \sparqlStar\ can be used to query {\rdfStarGraph}s.
To fully benefit from the extended data model, \sparqlStar\ adds new features that enable users to directly access metadata triples in queries. In particular, \sparqlStar\ introduces the possibility to bind {\rdfStarTriple}s to query variables; such a variable may then be used in a triple pattern in order to ask for matching metadata triples. Furthermore, as a shortcut, (recursively nested) triple patterns may be embedded directly in triple patterns (as demonstrated in Example~\ref{ex:QueryingReificationDoneRight}).

In the following,
Section~\ref{ssec:SPARQLstar:Base} introduces basic terminology and concepts.
Section~\ref{ssec:SPARQLstar:SyntaxAndSemantics} defines \sparqlStar\ based on P\'{e}rez et al.'s algebraic syntax of SPARQL~\cite{Perez09:SemanticsAndComplexityOfSPARQL}. Thereafter, Section~\ref{sec:W3CExt} provides the corresponding extension of the W3C specification of SPARQL~\cite{Harris13:SPARQL1_1Language}.

\subsection{Basic Terminology and Concepts} \label{ssec:SPARQLstar:Base}
The basic concepts for defining SPARQL queries and their semantics are triple patterns and solution mappings. A \emph{triple pattern} is a tuple $\symTP \in \bigl( \symAllVariables \cup \symAllIRIs \cup \symAllBNodes \cup \symAllLiterals \bigr) \times \bigl( \symAllVariables \cup \symAllIRIs \bigr) \times \bigl( \symAllVariables \cup \symAllIRIs \cup \symAllBNodes \cup \symAllLiterals \bigr)$ where~$\symAllVariables$ is a set of query variables that is disjoint from $\symAllIRIs$, $\symAllBNodes$, and $\symAllLiterals$, respectively. A \emph{solution mapping} is a partial mapping $\mu : \symAllVariables \rightarrow \bigl( \symAllIRIs \cup \symAllBNodes \cup \symAllLiterals \bigr)$. \sparqlStar\ extends these concepts by introducing a notion of \emph{{\tpStar}s} and \emph{{\solmapStar}s}.

\begin{definition} \label{def:TPStar}
	Let $\symAllStarTPs$ be an~(infinite) set of tuples that is defined recursively as follows:
	\begin{enumerate}
		\item $\symAllStarTPs$ includes all triple pattern, i.e., $\symAllStarTPs \supseteq \bigl( \symAllVariables \cup \symAllIRIs \cup \symAllBNodes \cup \symAllLiterals \bigr) \times \bigl( \symAllVariables \cup \symAllIRIs \bigr) \times \bigl( \symAllVariables \cup \symAllIRIs \cup \symAllBNodes \cup \symAllLiterals \bigr)$; and
		\item if $\symTP \in \symAllStarTPs$ and $\symTP' \in \symAllStarTPs$, then $\tuple{\symTP,p,o} \in \symAllStarTPs$, $\tuple{s,p,\symTP} \in \symAllStarTPs$ and $\tuple{\symTP,p,\symTP'} \in \symAllStarTPs$ for all $s \in ( \symAllVariables \cup \symAllIRIs \cup \symAllBNodes \cup \symAllLiterals )$, $p \in ( \symAllVariables \cup \symAllIRIs )$, and $o \in ( \symAllVariables \cup \symAllIRIs \cup \symAllBNodes \cup \symAllLiterals)$.
	\end{enumerate}
	Any tuple $\tuple{s,p,o} \in \symAllStarTPs$ is a \definedTerm{\tpStar}.
\end{definition}

\begin{definition} \label{def:SolMapStar}
	A \definedTerm{\solmapStar} is a partial mapping $\eta : \symAllVariables \rightarrow \bigl( \symAllStarTriples \cup \symAllIRIs \cup \symAllBNodes \cup \symAllLiterals \bigr)$.
\end{definition}

\noindent
Note that, in contrast to standard solution mappings that bind variables only to an IRI, a blank node, or a literal, a \solmapStar\ may bind a variable also to an \rdfStarTriple.


The following three definitions adapt the standard notions of \emph{compatibility} of solution mappings, \emph{merging} of solution mappings, and \emph{application} of solution mappings to {\solmapStar}s.

\begin{definition} \label{def:Compatibility}
	Two {\solmapStar}s $\eta$ and $\eta'$ are \definedTerm{compatible}, denoted by $\eta \compatible \eta'$,
		if, for every variable $?v \in \bigl( \fctDom{\eta} \cap \fctDom{\eta'} \bigr)$, $\eta(?v) = \eta'(?v)$.
\end{definition}

\newpage 

\begin{definition} \label{def:Merging}
	Let $\eta$ and $\eta'$ be two {\solmapStar}s that are compatible.
	The \definedTerm{merge} of $\eta$ and $\eta'$, denoted by $\eta \cup \eta'$, is a \solmapStar\ $\eta''$ that has the following three properties:
	\begin{enumerate}
		\item $\fctDom{\eta''} = \fctDom{\eta} \cup \fctDom{\eta'}$,
		\item $\eta''(?v) = \eta(?v)$ for all $?v \in \fctDom{\eta}$, and
		\item $\eta''(?v) = \eta'(?v)$ for all $?v \in \fctDom{\eta'} \setminus \fctDom{\eta}$.
	\end{enumerate}
\end{definition}

\begin{definition} \label{def:Application}
	The \definedTerm{application} of a \solmapStar\ $\eta$ to a \tpStar\ $\symTP$, denoted by $\eta[\symTP]$, is the \tpStar\ that can be obtained by replacing all variables in $\symTP$ according to $\eta$ (unbound variables must not be replaced).
\end{definition}

\subsection{(Algebraic) Syntax and Semantics of \sparqlStar} \label{ssec:SPARQLstar:SyntaxAndSemantics}

This section defines the semantics of the core fragment of \sparqlStar, which is represented based on an algebraic syntax that extends the algebraic SPARQL syntax introduced by P\'{e}rez et al.~\cite{Perez09:SemanticsAndComplexityOfSPARQL}.

\begin{definition} \label{def:Expression}
	A \definedTerm{\sparqlStar\ expression} is defined recursively as follows:
	\begin{enumerate}
		\item
			Any finite set of {\tpStar}s is a \sparqlStar\ expression, which is called a \emph{\bgpStar}\!.
		\item
			If $\symTP$ is a \tpStar\ and $?v$ is a variable, then $(\symTP \OpAS ?v)$ is a \sparqlStar\ expression.
		\item
			If $\symPattern_1$ and $\symPattern_2$ are \sparqlStar\ expressions and $R$ is a filter condition%
			\footnote{For a definition of the syntax of filter conditions refer to P\'{e}rez et al.'s work~\cite{Perez09:SemanticsAndComplexityOfSPARQL}.}%
			, then $(\symPattern_1 \OpAND \symPattern_2)$, $(\symPattern_1 \OpUNION \symPattern_2)$, $(\symPattern_1 \OpOPT \symPattern_2)$, and $(\symPattern_1 \OpFILTER R)$ are \sparqlStar\ expressions.
	\end{enumerate}
\end{definition}

\begin{example} \label{ex:Expression}
	The first query pattern of Example~\ref{ex:QueryingReificationDoneRight} can be represented as a \bgpStar\ $$\symPattern_\mathsf{ex} = \big\lbrace \tuple{?bob, \, \mathSmallRDF{foaf:name}, \, \mathSmallRDF{"Bob"}} , \tuple{ \tuple{?bob, \mathSmallRDF{foaf:age}, ?age}, \, \mathSmallRDF{dct:source}, ?src } \big\rbrace,$$ which consists of two {\tpStar}s. The second query pattern of Example~\ref{ex:QueryingReificationDoneRight} can be represented as a semantically equivalent \sparqlStar\ expression $\symPattern_\mathsf{ex2}$ that has the following form:
	$$\Bigl( \bigl(\tuple{?bob, \mathSmallRDF{foaf:age}, ?age} \OpAS ?t \bigr) \OpAND \big\lbrace \tuple{?bob, \, \mathSmallRDF{foaf:name}, \, \mathSmallRDF{"Bob"}} , \tuple{ ?t, \, \mathSmallRDF{dct:source}, ?src } \big\rbrace \Bigr) .$$
\end{example}

\noindent
The basis for defining the semantics of \sparqlStar\ is an algebra over multisets of {\solmapStar}s that resembles the standard SPARQL algebra (which is defined over multisets of ordinary solution mappings). Formally, a multiset {\solmapStar}s is a pair $M = \tuple{\Omega,\fctsymCard}$ where $\Omega$ is the underlying set (of {\solmapStar}s) and $\fctsymCard$ is the corresponding \emph{cardinality function}; i.e., $\fctsymCard : \Omega \rightarrow \lbrace 1,2, ... \, \rbrace$. Then, the \sparqlStar-specific algebra operators are defined as follows.

\newpage 

\begin{definition} \label{def:Algebra}
	Let $M_1 = \tuple{\Omega_1,\fctsymCard_1}$ and $M_2 = \tuple{\Omega_2,\fctsymCard_2}$ be multisets of {\solmapStar}s.
	\begin{itemize}
		\item
			The \definedTerm{join} of $M_1$ and $M_2$, denoted by $M_1 \Join M_2$, is a multiset of {\solmapStar}s $\tuple{\Omega,\fctsymCard}$ such that
			\begin{equation*}
				\Omega \definedAs \big\lbrace \eta_1 \cup \eta_2 \,\big|\, \eta_1 \in \Omega_1 \text{ and } \eta_2 \in \Omega_2 \text{ and } \eta_1 \compatible \eta_2 \, \big\rbrace
			\end{equation*}
			and, for every $\eta \in \Omega$,
			\begin{equation*}
				\fctCard{\eta} \definedAs \sum_{(\eta_1,\eta_2) \in \, \Omega^\eta} \fctsymCard_1(\eta_1) \cdot \fctsymCard_2(\eta_2),
			\end{equation*}
			where $\Omega^\eta = \big\lbrace (\eta_1,\eta_2) \in \Omega_1 \times \Omega_2 \,\big|\, \eta_1 \cup \eta_2 = \eta \big\rbrace$.

		\item
			The (multiset) \definedTerm{union} of $M_1$ and $M_2$, denoted by $M_1 \mcup M_2$, is a multiset of {\solmapStar}s $\tuple{\Omega,\fctsymCard}$ such that $\Omega \definedAs \Omega_1 \cup \Omega_2$ and, for every $\eta \in \Omega$,
			\begin{equation*}
				\fctCard{\eta} \definedAs \begin{cases}
					\fctsymCard_1(\eta) + \fctsymCard_2(\eta) & \text{if } \eta \in ( \Omega_1 \cup \Omega_2 ), \\
					\fctsymCard_1(\eta) & \text{if } \eta \in ( \Omega_1 \setminus \Omega_2 ), \\
					\fctsymCard_2(\eta) & \text{else}.
				\end{cases}
			\end{equation*}

		\item
			The (multiset) \definedTerm{difference} of $M_1$ and $M_2$, denoted by $M_1 \msetminus M_2$, is a multiset of {\solmapStar}s $\tuple{\Omega,\fctsymCard}$ such that $\Omega \definedAs \big\lbrace \eta \in \Omega_1 \,\big|\, \eta \incompatible \eta' \text{ for all } \eta' \in \Omega_2  \, \big\rbrace$ and $\fctCard{\eta} \definedAs \fctsymCard_1(\eta)$ for all $\eta \in \Omega$.

		\item
			The \definedTerm{left outer join} of $M_1$ and $M_2$, denoted by $M_1 \LJoin M_2$, is a multiset of {\solmapStar}s that is defined by:
			\begin{equation*}
				M_1 \LJoin M_2 \definedAs
				\bigl( M_1 \Join M_2 \bigr) \mcup \bigl( M_1 \msetminus M_2 \bigr) .
			\end{equation*}

		\item
			Given a filter condition $R$, the $R$-specific \definedTerm{selection} of $M_1$, denoted by $\sigma_R( M_1 )$, is a multiset of {\solmapStar}s $\tuple{\Omega,\fctsymCard}$ such that $\Omega \definedAs \big\lbrace \eta \in \Omega_1 \,\big|\, \eta \text{ satisfies } R \, \big\rbrace$ and $\fctCard{\eta} \definedAs \fctsymCard_1(\eta)$ for all $\eta \in \Omega$, where a \solmapStar\ $\eta$ \emph{satisfies} filter condition $R$ if any of the following holds:
			\begin{enumerate}
				\item $R$ is $\mathrm{bound}(?v)$ where $?v \in \symAllVariables$ and $?v \in \fctDom{\eta}$;
				\item $R$ is $?v = c$ where $?v \in \symAllVariables$ and $c \in \bigl( \symAllIRIs \cup \symAllLiterals \bigr)$, $?v \in \fctDom{\eta}$, and $\eta(?v) = c$;
				\item $R$ is $?x = \!?y$ where $?x,?y \in \symAllVariables$, $?x \in \fctDom{\eta}$, $?y \in \fctDom{\eta}$, and $\eta(?x) = \eta(?y)$;
				\item $R$ is $(\neg R')$ where $R'$ is a filter condition and $\eta$ does not satisfy $R'$;
				\item $R$ is $(R_1 \lor R_2)$ where $R_1$ and $R_2$ are filter conditions and $\eta$ satisfies $R_1$ or $R_2$; or
				\item $R$ is $(R_1 \land R_2)$ where $R_1$ and $R_2$ are filter conditions and $\eta$ satisfies both $R_1$ and $R_2$.
			\end{enumerate}

	\end{itemize}
\end{definition}

\noindent
Given these algebra operators, the semantics of any \sparqlStar\ expression is defined by the following evaluation function.

\newpage 

\begin{definition} \label{def:Evaluation}
	Let $\symPattern$ be a \sparqlStar\ expression and let $\symRDFstarGraph$ be an \rdfStarGraph. The \definedTerm{evaluation} of $\symPattern$ over $\symRDFstarGraph$, denoted by $\fctEval{\symPattern}{\symRDFstarGraph}$, is a multiset of {\solmapStar}s $\tuple{\Omega,\fctsymCard}$ that is defined recursively as follows:
	\begin{enumerate}
		\item
			If $\symPattern$ is a \bgpStar, then
			\begin{equation*}
				\Omega \definedAs \big\lbrace \eta \,\big|\, \eta\bigl[\rho[\symPattern]\bigr] \subseteq ( \fctTRefs{\symRDFstarGraph} \cup \symRDFstarGraph ) \text{ for some $\symPattern$-bnodes mapping } \rho \, \big\rbrace
			\end{equation*}
			and, for every $\eta \in \Omega$,
			\begin{equation*}
				\fctCard{\eta} \definedAs \Bigl| \big\lbrace \rho \,\big|\, \rho \text{ is a $\symPattern$-bnodes mapping such that } \eta\bigl[\rho[\symPattern]\bigr] \subseteq ( \fctTRefs{\symRDFstarGraph} \cup \symRDFstarGraph ) \big\rbrace \Bigr| ,
			\end{equation*}
			where a \emph{$\symPattern$-bnodes mapping} is a mapping $\rho : \mathrm{bn}(\symPattern) \rightarrow ( \symAllStarTriples \cup \symAllIRIs \cup \symAllBNodes \cup \symAllLiterals )$ and
			\begin{align*}
				\eta\bigl[\rho[\symPattern]\bigr] = \big\lbrace \eta[\symTP] \,\big|\, &\text{$\symTP$ is a \tpStar\ obtained by replacing all} \\[-1mm] &\text{blank nodes in some \tpStar\ $\symTP' \in \symPattern$ according to $\rho$ } \big\rbrace .
			\end{align*}

		\item
			If $\symPattern$ is $(\symTP \OpAS ?v)$, then
			\begin{equation*}
				\Omega \definedAs \big\lbrace \eta \,\big|\, \exists \eta' \in \Omega' : \eta' \compatible \eta \text{ and } \fctDom{\eta} \definedAs \bigl( \fctDom{\eta'} \cup \lbrace ?v \rbrace \bigr) \text{ and } \eta(?v) = \eta'[\symTP] \, \big\rbrace
			\end{equation*}
			and, for every $\eta \in \Omega$,
			\begin{equation*}
				\fctCard{\eta} = \sum_{\eta' \in \, \Omega'_\eta} \fctsymCard'(\eta') ,
			\end{equation*}
			where $\tuple{\Omega',\fctsymCard'} = \fctEval{\lbrace\symTP\rbrace}{\symRDFstarGraph}$ and $\Omega'_\eta = \lbrace \eta' \in \Omega' \,|\, \eta' \compatible \eta \rbrace$ for all $\eta \in \Omega$.

		\item
			If $\symPattern$ is $(\symPattern_1 \OpAND \symPattern_2)$, then $\tuple{\Omega,\fctsymCard} \definedAs \fctEval{\symPattern_1}{\symRDFstarGraph} \Join \fctEval{\symPattern_2}{\symRDFstarGraph}$.

		\item
			If $\symPattern$ is $(\symPattern_1 \OpUNION \symPattern_2)$, then $\tuple{\Omega,\fctsymCard} \definedAs \fctEval{\symPattern_1}{\symRDFstarGraph} \mcup \fctEval{\symPattern_2}{\symRDFstarGraph}$.

		\item
			If $\symPattern$ is $(\symPattern_1 \OpOPT \symPattern_2)$, then $\tuple{\Omega,\fctsymCard} \definedAs \fctEval{\symPattern_1}{\symRDFstarGraph} \LJoin \fctEval{\symPattern_2}{\symRDFstarGraph}$.

		\item
			If $\symPattern$ is $(\symPattern' \OpFILTER R)$, then $\tuple{\Omega,\fctsymCard} \definedAs \sigma_R\bigl( \fctEval{\symPattern'}{\symRDFstarGraph} \bigr)$.
	\end{enumerate}
\end{definition}

\begin{example} \label{ex:Evaluation}
	The evaluation of \sparqlStar\ expression $\symPattern_\mathsf{ex}$ (cf.~Example~\ref{ex:Expression}) over \rdfStarGraph\ $\symRDFstarGraph_\mathsf{ex}$~(cf.~Example~\ref{ex:RDFstarGraph}) consists of a single \solmapStar\ $\eta_1$, which has the following properties:
	\begin{enumerate}
		\item $\fctDom{\eta_1} = \lbrace ?bob, ?age, ?src \rbrace$,
		\item $\eta_1(?bob) = \mathSmallRDF{:bob}$,
		\item $\eta_1(?age) = \mathSmallRDF{23}$, and
		\item $\eta_1(?src) = \mathSmallRDF{http://example.net/homepage-listing.html}$.
	\end{enumerate}
	For the other expression from Example~\ref{ex:Expression} we obtain the same result: $\fctEval{\symPattern_\mathsf{ex2}}{\symRDFstarGraph_\mathsf{ex}} = \tuple{\Omega_\mathsf{ex2},\fctsymCard_\mathsf{ex2}}$ where $\Omega_\mathsf{ex2} = \lbrace \eta_1 \rbrace$ and $\fctsymCard_\mathsf{ex2}(\eta_1) = 1$.
\end{example}

\section{Extension of the W3C Specification of SPARQL} \label{sec:W3CExt}
After defining \sparqlStar\ based on an algebraic syntax, the remainder of this document defines a corresponding extension of the formalization of SPARQL~1.1 that is given by the W3C specification~\cite{Harris13:SPARQL1_1Language}.
This extension assumes that any mention of \textit{``RDF triple''} in the specification is understood as an \rdfStarTriple; similarly, \textit{``RDF graph''}, \textit{``triple pattern''}, \textit{``basic graph pattern''} (or \textit{``basic graph pattern''}, and \textit{``solution mapping''} are understood as \rdfStarGraph, \tpStar, \bgpStar, and \solmapStar, respectively. Furthermore, the understanding of a \textit{``property path pattern''} includes the possibility to use a \tpStar\
	as subject or object of such a pattern.

Section~\ref{ssec:W3CExt:Grammar} introduces the grammar of \sparqlStar\ as an extension of the grammar of SPARQL~1.1. Section~\ref{ssec:W3CExt:Translation} specifies how to support the extended grammar during the conversion of query strings into algebra expressions; this specification includes the introduction of a new algebra symbol which corresponds to \sparqlStar\ expressions of the form $(\symTP \OpAS ?v)$. Section~\ref{ssec:W3CExt:EvaluationSemantics} defines the evaluation semantics for the resulting algebra expressions. 

\subsection{Grammar} \label{ssec:W3CExt:Grammar}

This section specifies the \sparqlStar\ grammar as an extension of the standard SPARQL~1.1 grammar~\cite{Harris13:SPARQL1_1Language}. Elements of the grammar that are not specified explicitly in this section are defined as given in \cite[Section~19.8]{Harris13:SPARQL1_1Language}.

An \emph{embedded triple pattern} is a new syntax element that conforms to the following, new grammar~rules:
\begin{grammar}
EmbTP  ::=  '\textless\textless' \, VarOrBlankNodeOrIriOrLitOrEmbTP \, Verb \, VarOrBlankNodeOrIriOrLitOrEmbTP \, '\textgreater\textgreater' \par\vspace{1mm}
VarOrBlankNodeOrIriOrLitOrEmbTP ::=    Var $\vert$ \par
\hspace{60.5mm}                          BlankNode $\vert$ \par
\hspace{60.5mm}                          iri $\vert$ \par
\hspace{60.5mm}                          RDFLiteral $\vert$ \par
\hspace{60.5mm}                          NumericLiteral $\vert$ \par
\hspace{60.5mm}                          BooleanLiteral $\vert$ \par
\hspace{60.5mm}                          EmbTP \par
\end{grammar}

\noindent
As the given grammar rules indicate, an embedded triple pattern may contain other embedded triple patterns. 
%
Embedded triple patterns may be used in a query in the following two ways: \enumA~they are part of a BIND clause (which corresponds to \sparqlStar\ expressions of the form $(\symTP \OpAS ?v)$), or \enumB~they are embedded in a \tpStar\ or in a property path pattern. Both of these options are specified and discussed in the following.

SPARQL introduces the BIND clause for assigning the value of evaluating a given expression to a variable. To enable the use of embedded triple patterns in BIND clauses (instead of an expression), the following two extensions to the grammar are necessary. First, a new grammar rule is added:
\begin{grammar}
ExpressionOrEmbTP ::= Expression $\pmb{\vert}$ EmbTP
\end{grammar}
Second, the original grammar rule \texttt{[60]} is redefined as follows:%
\footnote{The adjusted part in which a redefined grammar rule differs from the original rule in \cite[Section~19.8]{Harris13:SPARQL1_1Language} is given in bold font.}
\begin{grammar}
Bind  ::=  'BIND' '(' \textbf{ExpressionOrEmbTP} 'AS' Var ')'
\end{grammar}

\noindent
An embedded triple pattern may not only be used in a BIND clause but it may also be embedded in a property path pattern or in a \tpStar. More precisely, in the extended syntax the subject or object of a property path pattern can be an embedded triple pattern (instead of an RDF term or a variable). Similarly, {\tpStar}s may have an embedded triple pattern in the subject position or in the object position (cf.~Definition~\ref{def:TPStar}). To this end, the grammar is extended with a new rule:

\begin{grammar}
VarOrTermOrEmbTP    ::=    Var $\vert$ GraphTerm $\vert$ EmbTP
\end{grammar}

\noindent
Moreover, the original grammar rules \texttt{[75]}, \texttt{[80]}, \texttt{[81]}, and \texttt{[105]} are redefined:

\begin{grammar}
TriplesSameSubject \hspace{5.5mm} ::=  \textbf{VarOrTermOrEmbTP} PropertyListNotEmpty $\vert$ TriplesNode PropertyList \par\vspace{1mm}
Object             \hspace{24mm} ::=    GraphNode $\pmb{\vert}$ \textbf{EmbTP} \par\vspace{1mm}
TriplesSameSubjectPath    ::=    \textbf{VarOrTermOrEmbTP} PropertyListPathNotEmpty $\vert$ \par\hspace{100mm} TriplesNodePath~PropertyListPath \par\vspace{1mm}
GraphNodePath \hspace{10.5mm}    ::= \textbf{VarOrTermOrEmbTP} $\vert$ TriplesNodePath
\end{grammar}

\subsection{Translation to the Algebra} \label{ssec:W3CExt:Translation}

Based on the SPARQL grammar the SPARQL specification \textit{``defines the process of converting graph patterns and solution modifiers in a SPARQL query string into a SPARQL algebra expression''}~\cite[Section~18.2]{Harris13:SPARQL1_1Language}. This process must be adjusted to consider the extended grammar introduced in the previous section.
	In the following, any step of the conversion process that requires adjustment is discussed.

\subsubsection{Variable Scope}
As a basis of the translation, the SPARQL specification introduces a notion of \emph{in-scope variables}~\cite[Section~18.2.1]{Harris13:SPARQL1_1Language}. To cover the new syntax elements introduced in Section~\ref{ssec:W3CExt:Grammar} this notion must be extended as follows.

\begin{itemize}
	\item A variable is in-scope of a \bgpStar\ $\symBGP$ if the variable occurs in $\symBGP$, which includes an occurrence in any embedded triple pattern in \bgpStar\ (independent of the level of nesting).
	\item A variable is in-scope of a property path pattern if the variable occurs in that pattern, which includes an occurrence in any embedded triple pattern in the pattern (independent of the level of nesting).
	\item A variable is in-scope of a BIND clause of the form \, \texttt{BIND ( T AS v )} \, (where \texttt{T} is an embedded triple pattern) if the variable is variable \texttt{v} or the variable occurs in the embedded triple pattern \texttt{T}. As for standard BIND clauses with expressions, variable \texttt{v} must \textit{``not \textnormal{[be]} in-scope from the preceeding elements in the group graph pattern in which \textnormal{[the BIND clause]} is used''}~\cite[Section~18.2.1]{Harris13:SPARQL1_1Language}. 
\end{itemize}

\subsubsection{Expand Syntax Forms}
The translation process starts with expanding \textit{``abbreviations for IRIs and triple patterns''}~\cite[Section~18.2.2.1]{Harris13:SPARQL1_1Language}. This step must be extended in two ways:

\begin{enumerate}
	\item Abbreviations for triple patterns with embedded triple patterns must be expanded as if each embedded triple pattern was a variable (or an RDF term).
		For instance, the following syntax expression \par
		\begin{footnotesize}\begin{verbatim}
			<<?c a rdfs:Class>> dct:source ?src ;
			                    prov:wasDerivedFrom <<?c a owl:Class>> .
		\end{verbatim}\end{footnotesize}
		must be expanded to
		\begin{footnotesize}\begin{verbatim}
			<<?c a rdfs:Class>> dct:source ?src .
			<<?c a rdfs:Class>> prov:wasDerivedFrom <<?c a owl:Class>> .
		\end{verbatim}\end{footnotesize}

	\item Abbreviations for IRIs in all embedded triple patterns must be expanded.
		For instance, the embedded triple pattern
		\begin{footnotesize}\begin{verbatim}
			<<?c a rdfs:Class>>
		\end{verbatim}\end{footnotesize}
		must be expanded to
		\begin{footnotesize}\begin{verbatim}
			<<?c <http://www.w3.org/1999/02/22-rdf-syntax-ns#type> <http://www.w3.org/2000/01/rdf-schema#Class>>>
		\end{verbatim}\end{footnotesize}
\end{enumerate}

\subsubsection{Translate Property Path Patterns} \label{sssec:W3CExt:Translation:Paths}
The translation of property path patterns (cf.~\cite[Section~18.2.2.4]{Harris13:SPARQL1_1Language}) has to be adjusted because the extended grammar allows for property path patterns whose subject or object is an embedded triple pattern (cf.~Section~\ref{ssec:W3CExt:Grammar}).

The translation as specified in the W3C specification distinguishes four cases. The first three of these cases do not require adjustment because they are taken care of either by recursion or by the adjusted translation of basic graph patterns (as defined in Section~\ref{sssec:W3CExt:Translation:BGPs} below). However, the fourth case must be adjusted as follows.

Let \, \texttt{X P Y} \, be a string that corresponds to the fourth case in \cite[Section~18.2.2.4]{Harris13:SPARQL1_1Language}. Given the grammar introduced in Section~\ref{ssec:W3CExt:Grammar}, \texttt{X} and \texttt{Y} may be an RDF term, a variable, or an embedded triple pattern, respectively (and \texttt{P} is a property path expression). The string \, \texttt{X P Y} \, is translated to the algebra expression \textsf{Path}( \texttt{X'}, \texttt{P}, \texttt{Y'}) where \texttt{X'} and  \texttt{Y'} are the result of calling a function named \texttt{Lift} for \texttt{X} and \texttt{Y}, respectively. For some input string \texttt{Z} (such as \texttt{X} or \texttt{Y}) that can be an RDF term, a variable, or an embedded triple pattern, the function \texttt{Lift} is defined as follows:
\begin{verbatim}
	If Z is an embedded triple pattern <<S,P,O>>
	    Return triple* pattern ( Lift(S), P, Lift(O) )
	Else
	    Return Z
	End
\end{verbatim}

\subsubsection{Translate Basic Graph Patterns} \label{sssec:W3CExt:Translation:BGPs}
After translating property path patterns, the translation process collects \textit{``any adjacent triple patterns [...] to form a basic graph pattern''}~\cite[Section~18.2.2.5]{Harris13:SPARQL1_1Language}. This step has to be adjusted because triple patterns in the extended syntax may have an embedded triple pattern in their subject position or in their object position (or in both). To ensure that every result of this step is a {\bgpStar}\!, before adding a triple pattern to its corresponding collection, its subject and object must be replaced by the result of calling function \texttt{Lift} (cf.~Section~\ref{sssec:W3CExt:Translation:Paths}) for the subject and the object,~respectively.

\subsubsection{Translate BIND Clauses with an Embedded Triple Pattern} \label{sssec:W3CExt:Translation:BIND}
The extended grammar in Section~\ref{ssec:W3CExt:Grammar} allows for BIND clauses with an embedded triple pattern.
The translation of such a BIND clause to a SPARQL algebra expression requires a new algebra~symbol:
\begin{itemize}
	\item \textsf{TR}( \tpStar, variable )
\end{itemize}

\noindent
Note that this symbol corresponds to \sparqlStar\ expressions of the form $(\symTP \OpAS ?v)$ (cf.~Definition~\ref{def:Expression}).

Then, any string of the form \,\, \texttt{BIND ( T AS v )} \,\, with \texttt{T} being an embedded triple pattern~(i.e., not a standard BIND expression) is translated to the algebra expression \textsf{TR}( \texttt{T'}, v ) where \texttt{T'} is the result of calling the aforementioned function \texttt{Lift} for \texttt{T}.

Notice, the translation of BIND clauses with an embedded triple pattern as defined in this section is used during the translation of group graph patterns that is specified in~\cite[Section~18.2.2.6]{Harris13:SPARQL1_1Language}. The case of BIND clauses with an embedded triple pattern is covered in this translation of group graph patterns by the last, ``catch all other'' \texttt{IF} statement (i.e., the \texttt{IF} statement with the condition \texttt{E is any other form}) and \emph{not} by the \texttt{IF} statement for BIND clauses with an~expression.

\subsection{Evaluation Semantics} \label{ssec:W3CExt:EvaluationSemantics}
The SPARQL specification defines a function \textit{``eval(D(G), algebra expression) as the evaluation of an algebra expression with respect to a dataset D having active graph G''}~\cite[Section~18.6]{Harris13:SPARQL1_1Language}. Recall that the active graph G in the context of \sparqlStar\ is an \rdfStarGraph, and so is any other graph in dataset D. The definition of function eval is recursive; the two base cases of this definition for \sparqlStar\ are given as follows:

\begin{itemize}
	\item For any \bgpStar\ $\symBGP$, $\mathrm{eval}\bigl( D(G), \symBGP \bigr) \definedAs \fctEval{\symBGP}{G}$ (where $\fctEval{\symBGP}{G}$ is the evaluation of $\symBGP$ over \rdfStarGraph~$G$ as per Definition~\ref{def:Evaluation}).
	\item For any algebra expression $E$ of the form \textsf{TR}( $\symTP, ?v$ ) where $\symTP$ is a \tpStar\ and $?v$ is a variable (as introduced in Section~\ref{sssec:W3CExt:Translation:BIND}), $\mathrm{eval}\bigl( D(G), E \bigr) \definedAs \fctEval{(\symTP \OpAS ?v)}{G}$ (where $\fctEval{(\symTP \OpAS ?v)}{G}$ is the evaluation of \sparqlStar\ expression $(\symTP \OpAS ?v)$ over \rdfStarGraph~$G$ as per Definition~\ref{def:Evaluation}).
\end{itemize}

\noindent
For any other algebra expression, the SPARQL specification defines algebra operators. These definitions can be extended naturally to operate over multisets of {\solmapStar}s (instead of ordinary solution mappings). Given this extension, the recursive steps of the definition of function eval for \sparqlStar\ are the same as in the SPARQL specification.

%
%

\bibliographystyle{alpha}
\bibliography{main}

\end{document}